\documentstyle[12pt]{article}
\input{epsf}

\makeatletter
\def\dash{-}
\def\@citex[#1]#2{%
\if@filesw\immediate\write\@auxout{\string\citation{#2}}\fi
  \def\@citea{}\@cite{\@for\@citeb:=#2\do
     {\ifx\dash\@citeb{--}\def\@citea{}\else
      \@citea\def\@citea{,\penalty\@m\ }\@ifundefined
       {b@\@citeb}{{\bf ?}\@warning
       {Citation `\@citeb' on page \thepage \space undefined}}\fi
\hbox{\csname b@\@citeb\endcsname}}}{#1}}
\newbox\tempboxa
\newdimen\captionboxsubcount 
\def\capsize#1{\captionboxsubcount=#1pt}
\newdimen\captionboxsub
\captionboxsub=\hsize \advance\captionboxsub by -\captionboxsubcount
\advance\captionboxsub by -\captionboxsubcount
\long\def\@makecaption#1#2{
 \setbox\@tempboxa\hbox{#1: #2}
 \ifdim \wd\@tempboxa >\captionboxsub 
\rightskip=\captionboxsubcount \leftskip=\captionboxsubcount #1: #2 
\else \hbox to\hsize{\hfil\box\@tempboxa\hfil} 
 \fi}
\makeatother
\capsize{30}

\newcommand{\bee}{\begin{equation}}
\newcommand{\ee}{\end{equation}}
\newcommand{\ba}{\begin{eqnarray}}
\newcommand{\ea}{\end{eqnarray}}
\newcommand{\bd}{\begin{displaymath}}
\newcommand{\ed}{\end{displaymath}}
\newcommand{\nonum}{\nonumber\\}

\newcommand{\Tr}{\hbox{Tr}}

\begin{document}

\begin{titlepage}

\begin{flushright}
\begin{minipage}{5cm}
\begin{flushleft}
SU--4240--639 \\
UNITU--THEP--10/1996 \\
hep-ph/9607412 \\
July, 1996
\end{flushleft}
\end{minipage}
\end{flushright}

\vskip2cm

\begin{center}
\Large\bf
Hidden Structure in a Lagrangian for Hyperfine Splitting of the Heavy
Baryons
\end{center}

\vspace{0.5cm}
\vfill

\begin{center}
\large
Masayasu {\sc Harada}$^{(a)}$\footnote{
Electronic address : {\tt mharada@npac.syr.edu}},
Asif {\sc Qamar}$^{(a)}$\footnote{
Electronic address : {\tt qamar@suhep.phy.syr.edu}},
Francesco {\sc Sannino}$^{(a,b)}$\footnote{
Electronic address : {\tt sannino@npac.syr.edu}},\\
Joseph {\sc Schechter}$^{(a)}$\footnote{
Electronic address : {\tt schechter@suhep.phy.syr.edu}}
and
Herbert {\sc Weigel}$^{(c)}$\footnote{
Electronic address : 
{\tt weigel@sunelc1.tphys.physik.uni-tuebingen.de}}
\end{center}

\vspace{0.5cm}

\begin{flushleft}
\it 
\qquad$^{(a)}$Department of Physics, Syracuse University, 
Syracuse, NY 13244-1130, USA
\\
\qquad$^{(b)}$%
Dipartimento di Scienze Fisiche \& Istituto
Nazionale di Fisica Nucleare \\
\qquad\qquad Mostra D'Oltremare Pad. 19,  80125 Napoli, Italy
\\
\qquad$^{(c)}$%
Institute for Theoretical Physics, 
T\"ubingen University \\
\qquad\qquad
Auf der Morgenstelle 14, D-72076 T\"ubingen, Germany
\end{flushleft}

\vfill

\begin{abstract}
We investigate the hyperfine splitting of the heavy baryons in the
bound-state approach.
We start with an ordinary relativistic Lagrangian which has been
extensively used to discuss finite mass corrections to the heavy
limit predictions.
It turns out that the dominant contribution arises from terms which do
not manifestly break the heavy spin symmetry.
The actual heavy spin violating terms are uncovered by carefully
performing a $1/M$ expansion of this Lagrangian.
\end{abstract}

\vspace{0.5cm}

\begin{flushleft}
\end{flushleft}

\end{titlepage}

\setcounter{footnote}{0}

\section{Introduction}

There has been a great deal of recent interest in studying 
heavy baryons in the bound-state 
picture~\cite{Callan-Klebanov,Blaizot-Rho-Scoccola}
together with heavy-quark spin symmetry~\cite{Eichten-Feinberg}.
This approach raises many fascinating questions which have been
explored by several groups~%
\nocite{Guralnik-Luke-Manohar,Jenkins-Manohar,Rho,Oh-Park,%
Gupta-Momen-Schechter-Subbaraman,Schechter-Subbaraman,%
Schechter-Subbaraman-Vaidya-Weigel}%
\cite{Guralnik-Luke-Manohar,-,Schechter-Subbaraman-Vaidya-Weigel}.

These models consist of a chiral Lagrangian for the light flavors 
and a Lagrangian, ${\cal L}_{\rm heavy}$, which contains the 
heavy-meson multiplet $H$. The simplest choice for the latter is~%
\cite{Schechter-Subbaraman:2}
\bee
{\cal L}_{\rm heavy}/M = i V_\mu \,
\Tr \left[H D_\mu \overline{H}\right]
+ id \, \Tr \left[ H \gamma_\mu \gamma_5 p_\mu \overline{H} \right]
\ ,
\label{heavy lag: leading}
\ee
where $V_\mu$ is the four-velocity of the heavy particle and
$D_\mu = \partial_\mu - i v_\mu$ is the covariant chiral derivative.
Furthermore
$v_\mu,p_\mu = (i/2) \left( \xi \partial_\mu \xi^{\dag}
\pm \xi^{\dag} \partial_\mu \xi \right)$, wherein 
$\xi = \exp\left(i\phi/F_\pi\right)$ is the non--linear 
representation of the light pseudoscalar mesons $\phi$. Finally $M$ 
is the heavy meson mass while $d$ is a heavy meson--light meson 
coupling constant. The light part of the Lagrangian allows for 
a soliton configuration $\xi_{\rm c}$. In the bound state approach 
the heavy baryon then emerges as a heavy meson bound state in the 
background of $\xi_{\rm c}$.
The predictions are very simple in the limit where both $N_{\rm C}$
and $M$ go to infinity.
For example, the binding energy of the heavy baryon~%
\cite{Guralnik-Luke-Manohar,Rho,Gupta-Momen-Schechter-Subbaraman}
is $(3/2)d\,F'(0)$ where $F'(0)$ is the slope of the soliton
profile at the origin.

An immediate question is how to estimate what happens when we
consider realistic values for 
$M$.
In general this requires the addition of many unknown terms to 
Eq.~(\ref{heavy lag: leading}).
A predictive model for finite $M$ corrections may be obtained by
constructing a Lagrangian ${\cal L}$~%
\cite{Schechter-Subbaraman:2,Yan-etal}
of
a heavy pseudoscalar meson
$P$ and a heavy vector meson 
$Q_\mu$:
\ba
{\cal L}(P,Q_\mu) &=&
- D_\mu P D_\mu {\overline P} - M^2 P {\overline P}
- \frac{1}{2} Q_{\mu\nu} \overline{Q}_{\mu\nu}
- {M^{\ast}}^2 Q_\mu {\overline Q}_\mu
\nonum
&&{}
+ 2 i M d 
  \left( P p_\mu {\overline Q}_\mu - Q_\mu p_\mu {\overline P} \right)
- i d' \epsilon_{\alpha\beta\mu\nu}
  \left(
    D_\alpha Q_\beta p_\mu {\overline Q}_\nu
    - Q_\alpha p_\beta D_\mu {\overline Q}_\nu
  \right)
\ .
\label{P Q lag}
\ea
This reproduces Eq.~(\ref{heavy lag: leading})
for large $M$ when $d'=d$ and $M^{\ast}=M$. We have used
$D_\mu \overline{P} = (\partial_\mu - i v_\mu)\overline{P}$
and
$\overline{Q}_{\mu\nu} = \left(
D_\mu {\overline Q}_\nu - D_\nu {\overline Q}_\mu \right)$.
This model in particular allows for manifest breaking of the heavy 
spin symmetry by choosing $M^{\ast}\neq M$ and/or $d'\neq d$.
The Lagrangian (\ref{P Q lag}) represents the starting point 
for computing physical quantities along the lines of the original 
bound state approach~\cite{Callan-Klebanov} to strangeness in the 
Skyrme model~\cite{Sk61,Ad83}. This requires the solutions to the 
equations of motion for $P$ and $Q_\mu$ in the soliton background. 
The calculation~\cite{Oh-Park,Schechter-Subbaraman-Vaidya-Weigel}
exhibits sizable corrections for finite $M$. In addition, recoil 
effects (finite $N_{\rm C}$) seem to be very important as well~%
\cite{Schechter-Subbaraman,Schechter-Subbaraman-Vaidya-Weigel}.
When both these effects are taken into account it becomes
difficult to fit the existing experimental data on the 
spectrum of the heavy baryons.  It was, however, noticed~%
\cite{Gupta-Momen-Schechter-Subbaraman,-,%
Schechter-Subbaraman-Vaidya-Weigel} that the inclusion of light 
vector mesons appreciably improves the situation.

In this note we will resolve an apparent puzzle which arises 
when calculating the corrections to the hyperfine splitting
using Eq.~(\ref{P Q lag}).

\section{An apparent puzzle}

First let us consider the calculation of the hyperfine splitting in
the heavy field approach.
This, of course, arises at first sub-leading order in $1/M$ and
violates the heavy spin symmetry.
Thus we must add to Eq.~(\ref{heavy lag: leading})
suitable heavy spin violating terms~\cite{Jenkins-Manohar}:
\bee
{\cal L}'_{\rm heavy}/M =
\frac{M-M^{\ast}}{8} \Tr 
\left[H \sigma_{\mu\nu} \overline{H} \sigma_{\mu\nu} \right]
- \frac{i}{2} (d-d') \Tr 
\left[ H p_\mu \overline{H} \gamma_\mu \gamma_5 \right]
+ \cdots 
\ .
\label{breaking terms}
\ee
The first term has no derivatives while the second term has one
derivative. The hyperfine splitting is related to a collective 
Lagrangian parameter (see section 4 for details) $\chi$
with a proportionality factor of the $\Delta$-$N$ mass difference:
\bee
m(\Sigma_Q^{\ast}) - m(\Sigma_Q) =
\left[ m(\Delta) - m(N) \right] \chi 
\ .
\label{mass diff}
\ee
(At present only $\Sigma_c$ is well established experimentally.)
For Eq.~(\ref{breaking terms}) we have
\bee
\chi = \frac{M^{\ast}-M}{4dF'(0)} 
+ \frac{d-d'}{4d}
\ .
\label{chi: 1}
\ee
The first term was obtained in Ref.~\cite{Jenkins-Manohar}
while the second seems to be new.
Notice that $(M^{\ast}-M)$ and $(d-d')$ behave as $1/M$.
These quantities are the same as the ones appearing in the ordinary 
field Lagrangian (\ref{P Q lag}).
It would thus seem that ${\cal L}'_{\rm heavy}$ in 
Eq.~(\ref{breaking terms}) neatly summarizes the heavy spin violation
in Eq.~(\ref{P Q lag}).

Now let us consider the calculation of $\chi$ from Eq.~(\ref{P Q lag})
directly based on exact numerical solution of the associated coupled
differential equations.
We content ourselves with the graphical presentation of some 
results\footnote{
For the Skyrme model parameters we use the experimental value of
$F_\pi$ and $e_{\rm Sk}=6.0$.
This results in a profile with $F'(0)=1.20$\,GeV.}
and relegate the details to a forthcoming 
publication~\cite{preparation}\footnote{%
Similar calculations were done in Ref.~\cite{Oh-Park}
but they did not consider the $M=M^{\ast}$, $d=d'$ case. }.
\begin{figure}[htbp]
\begin{center}
\ \epsfbox{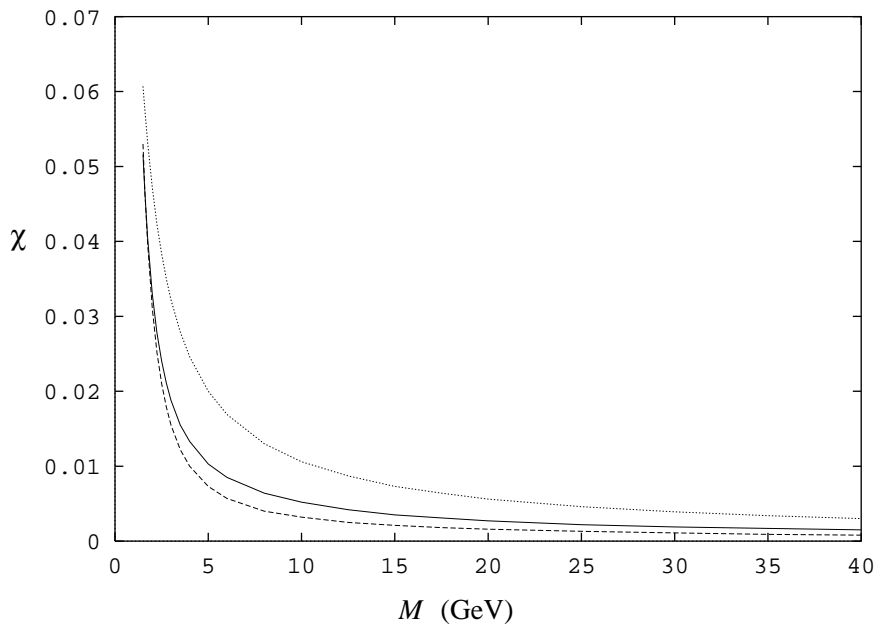} \\
\end{center}
\caption[]{
$\chi$ vs.~$M$ computed by numerical integration.
Solid line $M^{\ast}=M$, $d'=d$;
dotted line $M^{\ast}\neq M$, $d'=d$,
dashed line $M^{\ast}=M$, $d'\neq d$.
}
\label{fig: 1}
\end{figure}
Figure~\ref{fig: 1} shows $\chi$ plotted against $M$ for three cases:
i)~$M^{\ast}=M$, $d'=d=0.53$, 
ii)~$M^{\ast}-M\simeq(0.258\mbox{GeV})^2/M$ (a fit to experiment),
$d'=d=0.53$,
iii)~$M^{\ast}=M$, $d'-d=(0.0991\mbox{GeV})/M$ (an arbitrary choice
which sets the coupling constant splitting to be 10\% at the $D$ meson
mass).
We immediately notice that $\chi$ does not vanish when there is no
manifest heavy spin violation, i.e.,
$M=M^{\ast}$, $d=d'$.
In fact the dominant part of the contribution to $\chi$ for realistic
heavy meson masses is already present in this case.
By subtracting out this piece we note that the signs of the
contributions due to $M^{\ast}\neq M$ and $d'\neq d$ agree with those
predicted in Eq.~(\ref{chi: 1}).
It is interesting to note that all three curves in Fig.~\ref{fig: 1}
fall off as $1/M$ for $M\geq10$\,GeV.
But our main task is to understand the source of the puzzling non-zero
contribution in case i.
It is clear that the ordinary field Lagrangian (\ref{P Q lag})
must contain heavy spin violating pieces which are not manifest.
We will now explore this in detail by rewriting Eq.~(\ref{P Q lag}) 
in terms of the ``fluctuation field'' $H$ and expanding it in powers
of $1/M$.

\section{Expansion of Lagrangian}

Since the effects of $M\neq M^{\ast}$ and $d\neq d'$ were taken into
account in Eq.~(\ref{chi: 1}) it is sufficient to expand 
Eq.~(\ref{P Q lag}) with $M^{\ast}=M$ and $d'=d$.
To describe the heavy particle moving with four--velocity $V_\mu$,
we introduce the factorization 
\bee
P = e^{iMV\cdot x} P' \ , \qquad 
Q_\mu = e^{iMV\cdot x} \widetilde{Q}_\mu
\ .
\label{def: P Q}
\ee
$P'$ is the pseudoscalar ``fluctuation field''.
$\widetilde{Q}_\mu$ is not exactly the vector fluctuation field
since $V\cdot\widetilde{Q}$ is not constrained to be zero.
We therefore introduce the correct fluctuation field $Q'_\mu$ by
\bee
\widetilde{Q}_\mu = Q'_\mu - V_\mu V\cdot\widetilde{Q} 
\label{def: Q}
\ ,
\ee
which shows that $V\cdot Q'=0$.
Substituting Eqs.~(\ref{def: P Q}) and (\ref{def: Q}) into the
Lagrangian (\ref{P Q lag}) gives, in addition to the leading terms of
order $M$, the presently interesting terms of order $M^0$:
\ba
{\cal L} (P,Q) &=&
(\mbox{\rm order}\ M) +
P' D^2 \overline{P'} + Q'_\mu D^2 \overline{Q'}_\mu
- Q'_\mu D_\nu D_\mu \overline{Q'}_\nu
\nonumber\\
&&{}
+id \epsilon_{\alpha\beta\mu\nu}
\left(
 D_\alpha Q'_\beta p_\mu \overline{Q'}_\nu -
 Q'_\alpha p_\beta D_\mu \overline{Q'}_\nu
\right)
\nonumber\\
&&{}
+ M^2 V\cdot\widetilde{Q} V\cdot\overline{\widetilde{Q}}
- iM \left(
 D_\mu Q'_\mu V\cdot\overline{\widetilde{Q}} -
 V\cdot\widetilde{Q} D_\mu \overline{Q'}_\mu
\right)
\nonumber\\
&&{}
- 2iMd \left(
 P' V\cdot p V\cdot\overline{\widetilde{Q}} -
 V\cdot\widetilde{Q} V\cdot p \overline{P'}
\right)
+ \cdots
\ ,
\label{P Q lag 2}
\ea
where the three dots stand for terms of order $1/M$.
In contrast to the massless fields $P'$ and $Q'$,
$V\cdot\widetilde{Q}$ is seen to have the large mass $M$.
We thus integrate it out using the equation of motion
\bee
V\cdot\widetilde{Q} =
\frac{i}{M} D_\mu Q'_\mu +
\frac{2id}{M} P' V\cdot p 
\ .
\label{EOM for Q}
\ee
Substituting Eq.~(\ref{EOM for Q}) back into Eq.~(\ref{P Q lag 2})
gives
\ba
{\cal L} (P,Q) &=&
(\mbox{\rm order}\ M) +
P' D^2 \overline{P'} + Q'_\mu D^2 \overline{Q'}_\mu
- i Q'_\mu F_{\mu\nu}(v) \overline{Q'}_\nu
\nonumber\\
&&{}
- 2d \left(
 P' V\cdot p D_\mu \overline{Q'}_\mu +
 D_\mu Q'_\mu V\cdot p \overline{P'}
\right)
\nonumber\\
&&{}
+id \epsilon_{\alpha\beta\mu\nu}
\left(
 D_\alpha Q'_\beta p_\mu \overline{Q'}_\nu -
 Q'_\alpha p_\beta D_\mu \overline{Q'}_\nu
\right)
\nonumber\\
&&{}
- 4 d^2 P' \left(V\cdot p\right)^2 \overline{P'}
+ \cdots
\ ,
\label{P Q lag 3}
\ea
where $F_{\mu\nu}(v) = \partial_\mu v_\nu - \partial_\nu v_\mu 
- i[v_\mu,v_\nu]$.
In order to extract the heavy spin violating pieces it is convenient
to rewrite the order $M^0$
Lagrangian in terms of the heavy multiplet
field  
$H = \frac{i}{2}\left( 1 - i \gamma\cdot V \right) 
\left( \gamma_5 P' + \gamma \cdot Q' \right)$.
After some algebraic calculation we find
\ba
{\cal L} (H) &=&
{\cal L}_{\rm heavy}
- \frac{1}{2} \Tr \left[ H D^2 \overline{H} \right]
+ i \frac{1}{8} \Tr \left[
\left[ H , \gamma_\mu \gamma_\nu \right] F_{\mu\nu}(v) \overline{H}
\right]
\nonumber\\
&&{}
+ d
\Biggl[
\frac{1}{2} \Tr 
\left[ D_\mu H \gamma_\mu \gamma_5 (V \cdot p) \overline{H} \right]
- \frac{i}{4} \Tr \left[ 
\gamma \cdot D H \gamma_\mu \gamma_5 p_\mu \overline{H} 
\right]
\nonumber\\
&&\qquad{}
- \frac{i}{4} \Tr \left[
\gamma \cdot D H p_\mu \overline{H} \gamma_\mu \gamma_5
\right]
+ \frac{1}{8} \Tr \left[
\sigma_{\mu\nu} D_\alpha H \gamma_\alpha V\cdot p \gamma_5 
\sigma_{\mu\nu} \overline{H}
\right]
+ {\rm h.c.}
\Biggr]
\nonumber\\
&&{}
+ d^2 \left[
\frac{1}{2} \Tr \left[
H \left( V\cdot p \right)^2 \overline{H}
\right]
+ \frac{1}{4} \Tr \left[
\sigma_{\mu\nu} H \sigma_{\mu\nu} \left( V\cdot p \right)^2
\overline{H}
\right]
\right]
+\cdots
\ ,
\label{H lag}
\ea
where ${\cal L}_{\rm heavy}$ is given in 
Eq.~(\ref{heavy lag: leading}). At this stage we see 
that Eq.~(\ref{H lag}) actually contains pieces
which are not manifestly invariant under the heavy spin
transformations $H\rightarrow SH$, 
$\overline{H} \rightarrow \overline{H}S^{\dag}$. 
These pieces involve two derivatives.

\section{Hyperfine splitting}

We now sketch the computation of the portion of $\chi$ in 
Eq.~(\ref{mass diff}) which results from the ``hidden'' heavy spin
violation in Eq.~(\ref{P Q lag}) that has been made explicit in 
Eq.~(\ref{H lag}).
For this purpose one needs the collective Lagrangian of the quantum
variable $A(t)$ which is obtained after substituting
\bee
\xi(\mbox{\boldmath $x$},t) = A(t) \xi_{\rm c}(\mbox{\boldmath$x$})
A^{\dag}(t)
\ , \qquad
\overline{H}(\mbox{\boldmath $x$},t) 
= A(t) \overline{H_{\rm c}}(\mbox{\boldmath$x$})
\ ,
\ee
(where $\xi_{\rm c}(\mbox{\boldmath$x$})$ is the classical Skyrme
soliton and $\overline{H_{\rm c}}(\mbox{\boldmath$x$})$ is the heavy
meson bound-state wave function)
and integrating over $d^3x$.
The key dynamical variable is the ``angular velocity'' 
$\mbox{\boldmath $\Omega$}$ defined by 
$A^{\dag}\dot{A}=\frac{i}{2}
\mbox{\boldmath $\tau$}\cdot\mbox{\boldmath $\Omega$}$.
The bound-state wave function may be conveniently presented in the
rest frame where
\bee
\overline{H} \rightarrow
\left(
\begin{array}{cc}
0 & 0 \\
\overline{h}^a_{lh} & 0 \\
\end{array}
\right)
\ ,
\ee
with $a$, $l$, $h$ representing respectively the iso-spin, light spin
and heavy spin bivalent indices.
We write~\cite{Schechter-Subbaraman}
\bee
\overline{h}^a_{lh} = \frac{u(r)}{\sqrt{4\pi M}}
\left( \widehat{\mbox{\boldmath$x$}}
\cdot{\mbox{\boldmath $\tau$}}\right)_{ad}
\psi_{dl,h}
\ ,
\ee
where $u(r)$ is a radial wave function (assumed very sharply peaked
near $r=0$ for large $M$) and, to leading order in $M$, the ``angular
part'' of the ground state wave function is~\cite{%
Gupta-Momen-Schechter-Subbaraman,Schechter-Subbaraman}
\bee
\psi^{(1)}_{dl,h} = \frac{1}{\sqrt{2}} \epsilon_{dl}
\delta_{2h}
\label{psi 1}
\ .
\ee
The specific value of the index $h$ results from the choice
$G_3=G=1/2$ where $G$ is the ``grand spin''.
To next leading order in $M$ the ground state wave function receives a
heavy spin violating admixture of
\bee
\psi^{(2)}_{dl,h} = 
\sqrt{\frac{2}{3}} \delta_{d1} \delta_{l1} \delta_{h1} + 
\frac{1}{\sqrt{6}} \left(
 \delta_{d2} \delta_{l1} + \delta_{d1} \delta_{l2}
\right) \delta_{h2}
\ .
\label{psi 2}
\ee
Finally, the hyperfine splitting parameter $\chi$ is recognized by
expanding the collective Lagrangian~\cite{Callan-Klebanov},
in powers of $\mbox{\boldmath $\Omega$}$ and picking up the
linear piece $L_{\rm coll} = (\chi/2) \Omega_3 + \cdots$.
Noting that the $\Delta$--nucleon mass difference is given by 
the moment of inertia, which relates the angular velocity to 
the spin operator~\cite{Ad83}, this piece of the Lagrangian 
yields Eq. (\ref{mass diff}) after canonical quantization of 
the collective coordinates~\cite{Callan-Klebanov}.
There are two types of contribution to $\chi$.
The first type, from the heavy spin violating terms proportional to
$d$ in Eq.~(\ref{H lag}), corresponds to the evaluation of heavy spin
violating operators in the ground state (\ref{psi 1}).
The second type corresponds to the evaluation of heavy spin
conserving operators in the ground state which includes an admixture 
of Eq.~(\ref{psi 2}) due to the 
$\Tr\left[ \gamma_\mu \gamma_\nu H F_{\mu\nu}(v) \overline{H} \right]$
term in Eq.~(\ref{H lag}).
The net result for the ``hidden'' part of $\chi$ is
\bee
\chi = \frac{F'(0)}{4M}
\left( d - \frac{1}{2d} \right)
\ .
\label{chi: 2}
\ee
This equation is expected to hold for large $M$.
To this should be added the ``manifest'' part given in 
Eq.~(\ref{chi: 1}).

It is important to compare Eq.~(\ref{chi: 2}) with the result for
$\chi$ obtained by the exact numerical solution for the model based 
on Eq.~(\ref{P Q lag}). This is gotten as an integral over the 
properly normalized radial functions $\Phi(r),\ldots,\Psi_2(r)$ which 
appear in the P--wave solution of the bound state equation~%
\cite{Schechter-Subbaraman-Vaidya-Weigel}:
\begin{eqnarray}
P^{\dag}&=&A(t)\frac{\Phi(r)}{\sqrt{4\pi}}\
\hat{\mbox{\boldmath $r$}}\cdot
\mbox{\boldmath $\tau$}\rho e^{i\epsilon t}, \qquad
Q_4^{\dag}=\frac{i}{\sqrt{4\pi}}A(t)
\Psi_4(r)\rho e^{i\epsilon t},
\nonumber \\
Q_i^{\dag}&=&\frac{1}{\sqrt{4\pi}}A(t)\left[
i\Psi_1(r)\hat r_i+\frac{1}{2}\Psi_2(r)\epsilon_{ijk}
\hat r_j\tau_k\right]\rho e^{i\epsilon t}\ .
\label{pwavean}
\end{eqnarray}
The spinor $\rho$ labels the grand spin of the bound heavy
meson. The choice $G_3=+1/2$ corresponds to $\rho=(1,0)^{\dag}$.
The heavy limit bound state wave function in Eq.~(\ref{psi 1})
corresponds to the special choice
\bee
\Phi(r)\propto u(r) \ ,\quad 
\Psi_1(r)=-\Phi(r)\ ,\quad \Psi_2(r)=-2\Phi(r)
\quad {\rm and} \quad \Psi_4(r)=0
\ .
\label{pwhl}
\ee
The numerical
solution to the bound state equations exactly exhibits 
these relations for $M,M^{\ast}\to\infty$~%
\cite{Schechter-Subbaraman-Vaidya-Weigel}.

\begin{figure}[htbp]
\begin{center}
\ \epsfbox{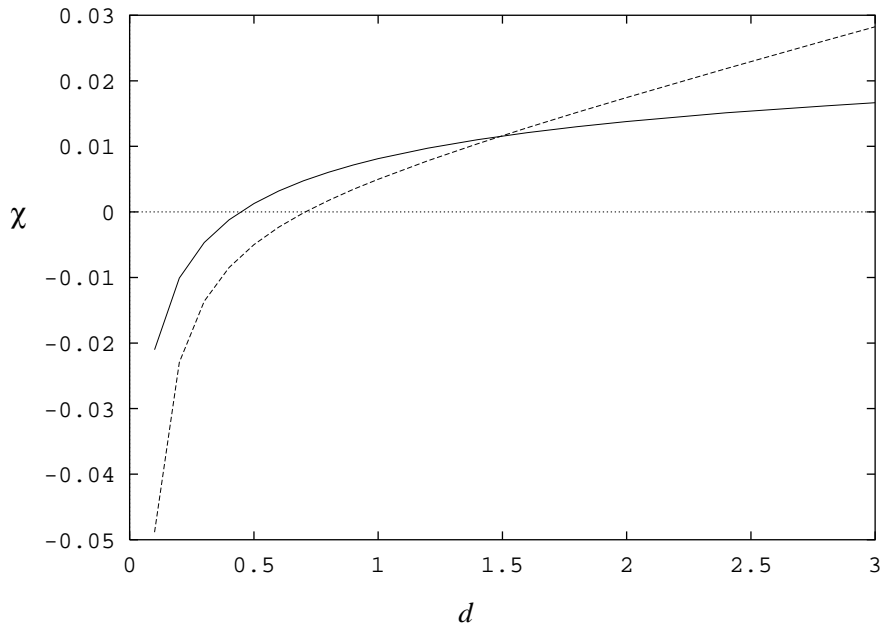}\\
\end{center}
\caption[]{
The $d$ dependence of $\chi$ for $M=M^{\ast}=30$\,GeV and $d=d'$.
Solid line is the exact numerical calculation.
Dashed line is the large $M$ perturbation formula given in 
Eq.~(\ref{chi: 2}).
}
\label{fig: 2}
\end{figure}

Equation~(\ref{chi: 2}) has an interesting $d$-dependence and vanishes
at $d=1/\sqrt{2}$, which actually is not too far from the experimental
value of this quantity.
In Fig.~\ref{fig: 2} we compare the $d$-dependence of the exact 
numerical calculation with the perturbative result of 
Eq.~(\ref{chi: 2}).
It is seen that the large $M$ perturbation approach works 
reasonably well and the gross structure of the hyperfine splitting 
is reproduced. For a detailed comparison of the two treatments 
it is important to note that for fixed $M=M^{\ast}$ the 
binding of the heavy meson increases with $d$. In particular 
this implies that the wave function is only reasonably 
localized for large enough $d$. As a strong localization is 
a basic feature of the perturbative approach it is easy to 
understand why this calculation does not yield the exact 
(numerical) result for small $d$. In fact, as $d$ increases 
the agreement expectedly improves. However, upon further increase 
of $d$ 
(at finite $M,M^{\ast}$),
the numerical solution to the bound state equations shows 
noticeable deviations from the heavy limit relations (\ref{pwhl}),
which causes the moderate differences at larger $d$.

\section{Discussion}

We have solved the apparent puzzle associated with the use of a model
Lagrangian containing ordinary fields for computing the hyperfine
splitting parameter $\chi$ by carefully expanding the Lagrangian in
powers of $1/M$.
The key point was the need to preserve the constraint 
$V\cdot Q\,'=0$ for the heavy vector fluctuation field.

Of course, such a model Lagrangian (which has been used in many
calculations) is not exactly QCD.
Nevertheless it seems reasonable since it automatically has the
correct relativistic kinematics and satisfies the heavy spin symmetry
at leading order.
We have seen (Eq.~(\ref{H lag})) that at next order in $1/M$, it
predicts the coefficients of many terms which otherwise would be
unspecified by heavy spin symmetry (even if reparameterization
invariance~\cite{Ki94} were taken into account).

It is amusing to note that these $1/M$ suppressed terms involve two
derivatives and are actually more important for the computation of
$\chi$ than the zero derivative term in Eq.~(\ref{chi: 1}).
This is readily understandable since the dynamical scale in this
calculation is the binding energy, 
$m(B)+m(N)-m(\Lambda_b)\simeq620$\,MeV which is rather large for
neglecting light vector mesons, higher derivatives etc.
[See, for example, Ref.~\cite{Harada-Sannino-Schechter}.]

We are regarding the Lagrangian (\ref{P Q lag}) as an illustrative
model rather than as a realistic one for comparison with experiment.
As indicated earlier it seems necessary to include, in addition
to finite $M$ corrections, the effects of light vector mesons as well
as nucleon recoil.
The discussion of $\chi$ in this more complicated model
and further details of the present calculation will
be given in a forthcoming publication~\cite{preparation}.

\section*{Acknowledgements}

One of us (HW) gratefully acknowledges the warm hospitality
extended to him during a visit at Syracuse University.

This work has been supported in part by the US DOE under contract
DE-FG-02-85ER 40231 and by the DFG under contracts We 1254/2--2 and
Re 856/2--2.

\end{document}